# Spaceborne, low-noise, single-photon detection for satellite-based quantum communications

**MENG YANG,**[1,2] **FEIHU XU,**[1,2] **JI-GANG REN,**[1,2] **JUAN YIN,**[1,2] **YANG LI,**[1,2] **YUAN CAO,**[1,2] **QI SHEN,**[1,2] **HAI-LIN YONG,**[1,2] **LIANG ZHANG,**[2,3] **SHENG-KAI LIAO,**[1,2,4] **JIAN-WEI PAN,**[1,2] AND **CHENG-ZHI PENG**[1,2,5]

[1]*Hefei National Laboratory for Physical Sciences at the Microscale and Department of Modern Physics, University of Science and Technology of China, Hefei 230026, China*
[2]*Chinese Academy of Sciences (CAS) Center for Excellence and Synergetic Innovation Center in Quantum Information and Quantum Physics, University of Science and Technology of China, Shanghai 201315, China*
[3]*Key Laboratory of Space Active Opto-Electronic Technology, Shanghai Institute of Technical Physics, Chinese Academy of Sciences, Shanghai 200083, China*
[4]*e-mail: skliao@ustc.edu.cn*
[5]*e-mail: pcz@ustc.edu.cn*

**Abstract:** Single-photon detectors (SPDs) play important roles in highly sensitive detection applications, such as fluorescence spectroscopy, remote sensing and ranging, deep space optical communications, elementary particle detection, and quantum communications. However, the adverse conditions in space, such as the increased radiation flux and thermal vacuum, severely limit their noise performances, reliability, and lifetime. Herein, we present the first example of spaceborne, low-noise, high reliability SPDs, based on commercial off-the-shelf (COTS) silicon avalanche photodiodes (APD). Based on the high noise-radiation sensitivity of silicon APD, we have developed special shielding structures, multistage cooling technologies, and configurable driver electronics that significantly improved the COTS APD reliability and mitigated the SPD noise-radiation sensitivity. This led to a reduction of the expected in-orbit radiation-induced dark count rate (DCR) from ~219 counts per second (cps) per day to ~0.76 cps/day. During a continuous period of continuous operations in orbit which spanned of 1029 days, the SPD DCR was maintained below 1000 cps, i.e., the actual in-orbit radiation-induced DCR increment rate was ~0.54 cps/day, i.e., two orders of magnitude lower than those evoked by previous technologies, while its photon detection efficiency was > 45%. Our spaceborne, low-noise SPDs established a feasible satellite-based up-link quantum communication that was validated on the quantum experiment science satellite platform. Moreover, our SPDs open new windows of opportunities for space research and applications in deep-space optical communications, single-photon laser ranging, as well as for testing the fundamental principles of physics in space.

## 1. Introduction

Single-photon detectors (SPDs) have been extensively used as one of the key components in extremely weak light detectionn applications over the past two decades. These include light detection and ranging [1-3], laser time transfer [4, 5], scintillation detection of elementary particles [6], deep-space optical communications [7, 8], and quantum communications [9-16]. Among these applications, quantum communications constitute the most rigorous application that demands increased SPD detection efficiency (DE) and low-dark count rate (DCR), given that the optical signal cannot be amplified like classical light signals according to the quantum noncloning principle [17] to guarantee the high fidelity of the single photon during its transmission. In addition to the establishment of a feasible large-scale quantum communication, increased DE, and low DCR, SPDs can effectively improve and extend existing application performances. For example, use of the SPD to replace the Lidar's light power detector can significantly reduce the dependence of high-power lasers and optical amplifiers [18, 19], and can efficiently simplify the system design and reduce power consumption.

At present, the available SPDs mainly include photomultiplier tubes (PMTs), avalanche photodiodes (APDs), and superconducting nanowire single-photon detectors (SNSPDs). In satellite-based quantum communication which is the primary interest of this work, an extensively used wavelength is near 800 nm. In this wavelength, PMTs suffer from low-detection efficiency (below 10%) [20, 21], and SNSPDs require heavy and bulky cryocoolers. By contrast, low-cost and compact silicon-based APDs have an ~65% detection efficiency near the 800 nm wavelength, and are more suited for our space-based missions [22]. Given the use of state-of-the-art, silicon-based APDs, the Execelitas's commercial off-the-shelf (COTS) super-low K factor (SLiK) silicon APD can achieve a higher DE with a special electric field distribution [23] compared with other silicon APD structures, such as reach-through structures. However, this also led to higher radiation-noise sensitivity [24]. However, given the adverse conditions in space, the silicon-APD-based SPD performances, especially the noise performance and reliability would face unprecedented challenges, as evidenced in previous pioneering work and space applications, such as in the National Aeronautics and Space Administration's (NASA's) ice, cloud, and land elevation satellite (ICESat) [1-3, 25] and in Chinese navigation satellites [5]. For example, in NASA's ICESat, eight lab-assembled, silicon-based APD SPDs were intended for cloud and aerosol backscattering measurements in the geoscience laser altimeter system (GLAS). Even though they were carefully designed and implemented, three of them failed during the GLAS system level tests prior to the ICESat launch [26]. Owing to space radiation, their radiation-induced DCR increased ~60 cps per day per SPD during orbit [3], which is unacceptable for quantum communications.

As the main promoter of the spaceborne, low-noise SPD, the quantum experiment science satellite (QUESS), i.e., Micius, is the world's first satellite designed to carry out quantum experiments [27, 28] at the space scale. One of its targets is to conduct ground-satellite quantum teleportation [15] that constitutes the primary interest of our work. The ground–satellite quantum teleportation is the first application of the spaceborne, low-noise SPD, receives uplink configurations, and places the complicated multiphoton setup on the ground. Compared to downlink configurations, two substantial challenges for uplink channels are the atmospheric turbulence at the beginning of the transmission path, which would cause a) increased link attenuation owing to beam wandering and broadening, and b) in-orbit receiver SPD radiation-induced DCR increases, thereby demanding SPDs with increased DEs and low DCR.

To evaluate the SLiK silicon APD sensitivity to radiation damage and in-orbit performance, we have exposed the APDs in the proton flux produced in the laboratory. According to the experimental results, the in-orbit radiation-induced DCR increment rate of the SLiK silicon APD would be ~219 cps/day. To mitigate the APD radiation-induced DCR increment rate and guarantee the SPD reliability, we have developed special radiation shield structures, multistage cooling technologies, and special driver electronics for the spaceborne, low-noise SPD. Based on these measures, the expected SPD in-orbit DCR increment rate would be ~0.76 cps/day, thus satisfying the satellite-based quantum communication applications.

Before launch, a series of space environmental tests were carried out to verify the SPD performances and reliabilities, and the results were proven satisfactory. The SPDs were then launched into orbit onboard the QUESS on 16 August 2016. During the past two years of in-orbit operations, we observed an ultralow DCR increase of ~0.54 cps/day by cooling the APDs to ~ -50 ℃, and by adjusting their excess voltages. This DCR increase was two orders of magnitude lower than previous results [3]. As the first example of spaceborne low-noise SPDs, they were used in the first demonstration of a ground–satellite quantum teleportation experiment [15], during which the DCRs of the SPDs remained below 150 cps. Furthermore, we evaluated numerically the importance of spaceborne, low-noise, Si APD SPDs in ground-to-satellite quantum key distributions (QKDs) [29-34]. The results showed that our SPDs could support ground-to-satellite QKDs for 4.6 years compared to conventional detectors which can only last a month.

## 2. Challenges and solutions

### 2.1 Radiation damage and simulations

Space radiation comprises mainly solar energetic particle events, galactic cosmic radiation, and trapped radiation belts. For a low Earth orbit (LEO), the major radiation sources are trapped radiation belts which consist of trapped electrons and protons in the South Atlantic Anomaly, and solar protons at the poles of the Earth's magnetosphere [35-37]. Trapped electrons can be shielded effectively to negligible levels with aluminum with a thickness of ~3 mm [24]. However, the radiation of protons has two major effects on the silicon lattice in silicon-based APDs, namely, ionizing dose damage and accumulated displacement damage [38]. For silicon APDs, the accumulated displacement damage is the major factor that increases the bulk leakage current appreciably or the DCR when the APD is operated in the Geiger mode.

The displacement damage creates different types of silicon lattice defects in the depletion region of APD, whereby the primary initial defects include vacancies, and interstitials. Accordingly, complex defects, or defect centers, are then formed with various numbers of vacancies and interstitials, such as Frenkel pairs, divacancies, vacancy–phosphorus pairs, and defect clusters [39]. These defect centers constitute new electron–hole pair generation centers, influence the properties of existing generation centers [40], and subsequently increase the DCR of the APD. The concentration of the defects depends only on the radiation particle nonionizing energy loss (NIEL, total energy that goes into displacements) and not on the type and initial energy of the particle [41, 42]. Based on this fact, it is practical to use mono-energetic particles, such as 50 MeV protons, to scale and simulate the displacement damage of different particles. Therefore, we used 50 MeV protons, which can penetrate the front glass window (the window in front of the APD was 1.0 mm thick and was made of Borosilicate glass which had a density of 2.6 g/cm$^3$ ) of APD package with slight spectra spreading to estimate and simulate the displacement damage to the SLiK silicon APD produced by the radiation particles in the in-orbit environment. In the case of the QUESS, its orbit features a circular Sun-synchronous orbit of approximately 500 km and an inclination of 97.4 °. Based on the space environment information system (SPENVIS) [43], we used the AP8-min and AE8-min models to evaluate the total displacement damage dose, which was equal to ~ $1.60 \times 10^9$ protons/cm$^2$ (this is equivalent to 50 MeV protons behind the 3 mm thick aluminum of the satellite case) for durations of two years, or ~ $2.19 \times 10^6$ protons/cm$^2$ per day. In the case of the APD operating in Geiger mode, its DCR can be expressed as [44],

$$\text{DCR} = S \cdot \int_{W_1}^{W_1+W_2} P_{\text{pair}}(x) \cdot G_{\text{TOT}} dx \qquad [\text{Eq. (1)}]$$

where $W_1$ and $W_2$ are the position and length of the depletion in the APD, $P_{pair}$ is the total probability of induction of an avalanche breakdown for an electron or a hole, and $G_{TOT}$ is the total carrier generation rate. According to the Shockley–Read–Hall theory, and following the counting of all carrier generation centers induced by the material's natural or radiation-induced defects, the total carrier generation rate, $G_{TOT}$, can be expressed as,

$$G_{TOT} = \sum_{d=1}^{D} \frac{n_i}{\tau_n \exp\left[\frac{-(E_{T,d} - E_i)}{KT}\right] + \tau_p \exp\left[\frac{-(E_{T,d} - E_i)}{KT}\right]} \qquad [\text{Eq. (2)}]$$

where $D$ is the number of defects in the depletion area, which increases almost linearly with the displacement damage dose or the in-orbit time, $E_{T,d}$ is the $d_{th}$ carrier generation center energy level, $E_i$ is the intrinsic Fermi energy level, and $n_i$ is the intrinsic carrier concentration. Additionally, $\tau_n$ and $\tau_P$ are the lifetimes of the electron and hole, respectively, $K$ is the Boltzmann constant, and $T$ is the absolute temperature.

As the radiation dose or in-orbit time increase, the number of defects in the APD depletion area increases. Correspondingly, the APD DCR increases linearly as a function of the displacement damage dose or the in-orbit time. However, owing to the different electron–hole pair generation rate for different carrier generation centers, the variations of the DCR increment rate as a function of the displacement damage dose or the in-orbit time would temporarily deviate from linearity at small dose spans or short in-orbit time spans (e.g., days or months). However, it would be a very difficult task to derive the APD macroscopic properties, such as the noise performance, from the characteristics of the individual microscopic carrier generation center, which needs to know the carrier generation center energy level in the bandgap, the carrier generation center concentrations associated with each level, the capture and emission probabilities for electrons and holes for each level, etc. In the practical analysis and calculation of the APD noise performance, a more direct and easier mode is used in the APD radiation damage [45-47], namely the DCR increment $\Delta DCR$. Owing to radiation damages, this variable can be expressed as,

$$\frac{\Delta DCR}{V} = \frac{n_i \phi}{2K_{gn}} \qquad [\text{Eq. (3)}]$$

where $V$ is the APD depletion region volume, $n_i$ is the intrinsic carrier density, $\phi$ is the radiation dose, and $K_{gn}$ is the damage coefficient for the material type in the depletion region. In the case of the SLiK APD, $V$, $n_i$, and $K_{gn}$, are constants. The DCR increment is linearly proportional to the radiation dose.

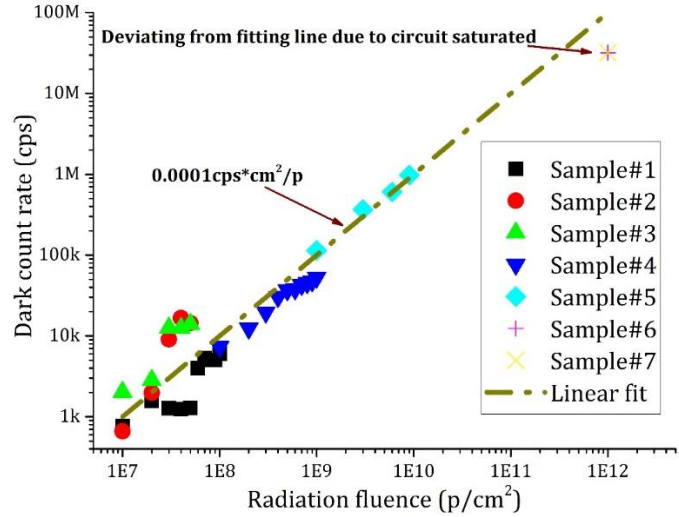

Fig. 1. Dark count rate (DCR) versus radiation fluence. During radiation, sample #2 was cooled to -35 °C and was biased with an excess voltage of 17 V, while samples #3 and #7 were unbiased at 25 °C, and the other samples were cooled to -12 °C and biased with an excess voltage of 17 V. For DCR measurements, all samples were cooled to -12 °C and biased with an excess voltage of 17 V.

To evaluate the APD performance degradation in the presence of a radiation environment, we have performed a series of radiation tests on the ground to evaluate how the displacement damage affects the macroscopic properties of the APDs. Seven samples based on Excelitas SLiK APDs and labeled #1–7 were radiated with a proton flux of 50 MeV. Fig. 1 shows the DCRs during the radiation test, which increase linearly as a function of radiation fluence in agreement with previous findings [1]. For an operating temperature of -12 °C and for an excess voltage of 17 V, the DCR sensitivity is $\sim 1.00 \times 10^{-4} \, \text{cps} \cdot \text{cm}^2 / \text{proton}$ (50 MeV). Combined with the estimated in-orbit radiation dose, the DCR increase is ~219 cps/day, which is too high for several space applications, including quantum communications [9-16]. In addition to DCRs, we also measured the APD breakdown voltage, responsivity, and dark current. No significant changes were observed for the responsivity and dark current, whereas the breakdown voltage presented a measurable increment up to ~3 V when the radiation fluence reached $1.00 \times 10^{12} \, \text{protons} / \text{cm}^2$. However, this fluence value is far beyond the total radiation dose during two years of QUESS operations.

## 2.2 Shield and thermal design

Plainly, an effective direct way to reduce the radiation-induced APD DCR increment rate is to reduce the radiation dose to APD, such as the selection of low orbits, small-orbit inclination angles, and the addition of protection shield layers. For a given experimental mission, the satellite orbit parameters are basically unmodifiable. Theoretically, radiation can be fully shielded with a shield layer with a sufficient thickness, but it needs extremely thick shields owing to secondary particle effects [48, 49]. Based on the use of the SPENVIS tools, we had evaluated the shielding effects of the shield layers with different aluminum thicknesses, as shown as Fig. 2. As the shield thickness increases, the secondary particles dominate in the APD equivalent radiation dose, and the equivalent 50.0 MeV proton fluence remains almost unchanged after the Al shield thickness reaches 200 mm. According to the previous estimation, the equivalent 50.0 MeV proton fluence should be lower than $7.30 \times 10^6 \, \text{protons}/\text{cm}^2$, while the in-orbit radiation of the SPD induced a DCR increment which was less than 1 cps/day. This required an Al shield thickness > 400 mm. Correspondingly, the quality of the shielding layer was ~723 kg, which is unaffordable for most satellites, especially for small satellites.

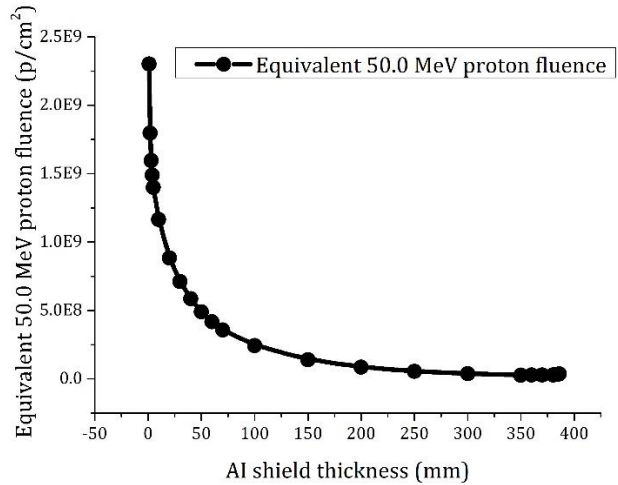

Fig. 2. Equivalent 50.0 MeV proton fluence (p/cm$^2$) versus spherical Al shield radius.

Taking into account the weight and volume limitation of the satellite, we have traded off the shield performance and the volume, weight, and thermal capacity. Furthermore, an aluminum shielding layer with a thickness of 12 mm—whose specific heat capacity was ~4 times of tantalum—and a tantalum shielding layer with a thickness of 4 mm were added around the APD, the total quality was ~5 kg, and the equivalent aluminum thickness was ~22 mm. This reduced the radiation dose damage by ~2.5 times. In addition, to achieve the APD deep cooling, the radiation shield layer was thermally isolated and cooled to approximately -15 ℃ by radiation cooling. The shielding and installation of the spaceborne low-noise Si APD single photon detectors, namely, SPD #1 and SPD #2, as shown in Fig. 3.

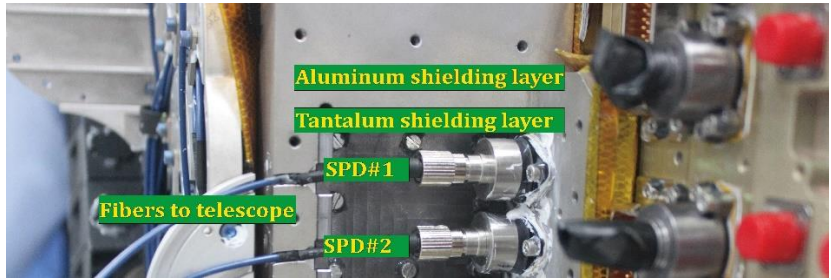

Fig. 3. Shielding and installation of the spaceborne, low-noise, Si APD, single photon detectors.

According to Eq. (2), the carrier generation rate of the carrier generation center decreased exponentially as a function of temperature. Consequently, the APD DCR can be effectively reduced by cooling the APD to lower temperatures, which is consistent with previous findings [50, 51]. Furthermore, we performed a series of tests on our APDs at different operating temperatures, and excess voltages to validate the effectiveness of cooling to mitigate the radiation-induced DCR.

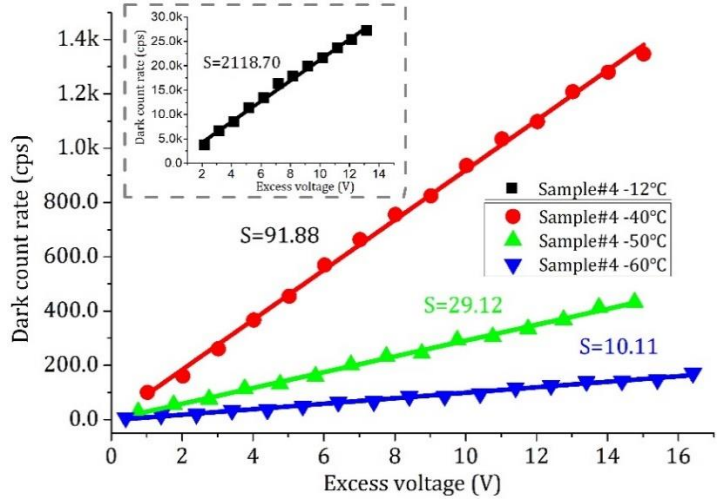

Fig. 4. Deep cooling effect on DCR of sample #4. The results for the other six samples are the same. S represents the slope for the DCR versus excess voltage plots at different temperatures.

Fig. 4 shows the test results for one irradiated sample, illustrating that the DRC increases exponentially as a function of temperature and linearly as a function of the excess voltage. In addition, the slope S for DCR versus the excess voltage also increases exponentially as a function of the operating temperature T. For sample #4, the slope can be expressed as,

$$S = \alpha \exp(\beta T) \tag{4}$$

We obtained $\alpha = 6807.53$ and $\beta = -0.09757$ from the testing data. According to Eq. (4), at the same excess voltage, the DCR decreased by ~2.65 times for a temperature drop of 10 °C. This demonstrated the effectiveness of deep cooling to mitigate the radiation damage. By further cooling the APD operating temperature from -12 °C to -60 °C, the DCR decreased by ~108 times. Combination of the shield and deep cooling yielded the expected in-orbit DCR increment of ~0.76 cps/day.

*2.3 Driving electronics and tests*

In our implementation, each SPD contained an Excelitas SLiK APD in a TO-can, with an active diameter of 180 μm and ~60% photon detection efficiency at a wavelength of 800 nm, a two-stage thermal electrical cooler (TEC), and a thermistor. The two-stage TEC and the thermistor were used to realize closed-loop control of the temperature. The package was gas-sealed to reduce the damage from other external factors, such as water vapor and extreme temperature. The driving circuit of each SPD, as shown in Fig. 5, included a robust passive quenching circuit, a high-voltage (HV) manager, an operating temperature controller, and a signal discrimination circuit.

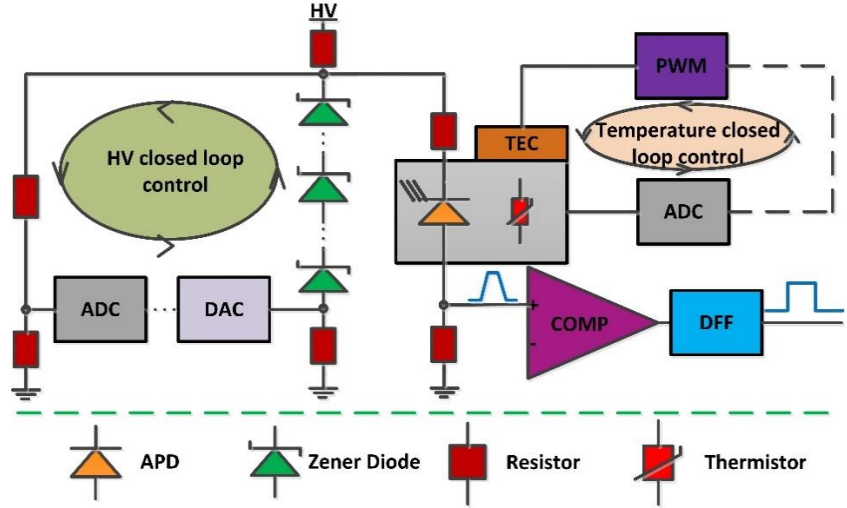

Fig. 5. Schematic of the driving circuit for single-photon detectors (SPDs) (ADC, analog-to-digital-converter; DAC, digital-to-analog-converter; COMP, comparator; DFF, D Flip-flop; PWM, pulse-width modulation; TEC, thermal electric cooler, HV, high voltage. The ADCs, DACs, and PWMs, are connected to a field-programmable gate array (FPGA) in which the temperature and the high-voltage (HV) closed loop control were carried out.

According to Eq. (1), the APD DCR was decided by the carrier generation rate and the total probability, $P_{pair}$. The avalanche breakdown increased linearly as a function of the excess voltage (the voltage above the APD breakdown voltage) because the electric field in the APD depletion region was basically uniform based on SLiK structure, as shown in Fig. 4. The APD itself DCR can be expressed as,

$$d_s = SVe \qquad [Eq. (5)]$$

where S is the slope of the plot of the DCR versus the excess voltage which depends on the operating temperature and radiation dose, and $Ve$ is the excess voltage. Conversely, the APD detection efficiency was decided by the quantum efficiency and the photon-induced carrier breakdown probability. The APD detection efficiency is a constant after the APD was produced. By contrast, the photon-induced carrier breakdown probability increased exponentially with the excess voltage [23]. Thus, the APD detection efficiency can be approximately expressed as,

$$\eta_d = \eta(1 - \exp(-Ve/Vc)) \qquad [Eq. (6)]$$

in which $\eta$ is the APD quantum efficiency and equals ~0.8, $V_c$ is the APD characteristic voltage which depends on the depletion layer thickness and the weighted average of the ratio of the ionization coefficient of holes to that of electrons. In the case of the SLiK APD, $V_c$ was ~9.1 V, according to our test results. Based on Eqs. (5–6), we can further optimize the system performance through the adjustment of the excess APD voltage. For example, in the uplink ground–satellite experiment [15], the teleportation fidelity can be expressed as,

$$F = \frac{\eta_l \eta_d F_s F_l + 1.795 R_d t_w}{\eta_l \eta_d + 3.59 R_d t_w} \qquad [Eq. (7)]$$

where $\eta_l$ is the is the link efficiency from the ground photon source to the input of APD, and is approximately equal to -45 dB, and $\eta_d$ is the APD detection efficiency. $F_s$ and $F_l$ are the photon source and link fidelities, which are equal to ~0.8 and ~0.99, respectively. $R_d$ is the

APD dark count rate, including the link background photon count rate which is ~150 cps, the thermal noise dark count and the radiation-induced dark count. Additionally, $t_w$ is the time window, which is ~1 ns. According to Eqs. (5–7), we can obtain the teleportation fidelity at different excess voltages and radiation damage levels, as shown in Fig. 6.

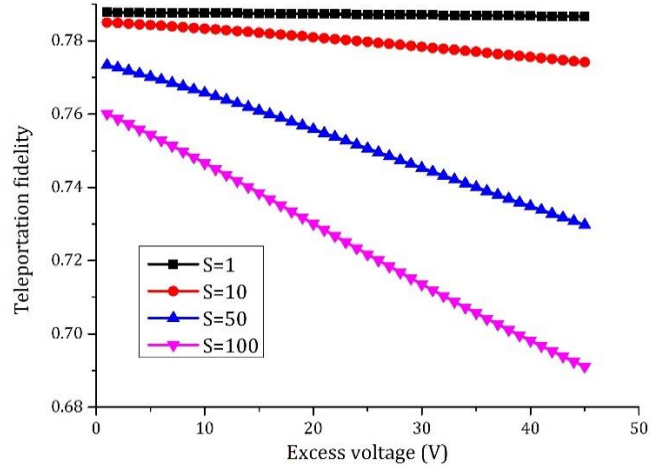

Fig. 6. Teleportation fidelity at different excess voltages and excess DCR voltages slopes.

Lowering the APD excess voltage can improve the teleportation fidelity, especially when the excess DCR voltage slope S is very large owing to radiation damage. Of course, lowering the APD excess voltage would reduce the APD detection efficiency. Accordingly, it will take more time to obtain an adequate number of effective detection events. Based on this fact, we have designed an adjusted HV manager with a broad adjustable range. In the HV manager, a coarse voltage of ~300 V is generated using voltage regulator diodes. The high voltage is further tuned through a digital-to-analog converter (DAC) for APD at different operating temperatures and detection efficiencies, as shown in Fig. 5. In addition, the temperature drift of the voltage regulator diodes was compensated with the use of a closed-loop controller.

For deep-cooling APD, multistage cooling technologies were introduced, including a stage radiation cooling and a two-stage TEC. The former was implemented with a heat pipe, the shield layer was cooled to $\sim 15°C$, and the APD was installed on the shield layer. To achieve additional cooling, the APD and fine control of the APD operating temperature, a digital proportional, integral, derivative (PID) closed-loop controller based on a two-stage TEC and a thermistor, were implemented. The value of the operating temperature was adjusted according the APD's DCR in a radiation environment to improve SPD reliability and save satellite energy.

In addition, thermal annealing also partially reduced the radiation-induced DCR [24, 52]. We annealed radiated samples #2 and #3 in an oven at 50 ℃ and 80 ℃ for 15 h and 265 h, respectively. The results illustrated that thermal annealing could indeed reduce the radiation-induced DCR. To extend the orbital lifetimes of the APDs, it is valuable to reserve their thermal annealing abilities. Herein, we used an H-bridge pulse-width modulation (PWM) driver to drive the two-stage TEC, which could heat or cool the APD. Accordingly, the APD could operate between approximately +80 ℃ and -60 ℃, as required. Note that laser annealing is another way to mitigate radiation-induced damages, as recently reported in [53], but it needs high-power laser sources.

For the Excelitas SLiK APD, the after-pulsing probability decreased exponentially with time and its tail attenuated to very low values after a few hundred nanoseconds [54]. A passive quenching circuit [55, 56] with a dead time of a few microseconds was used to drive the APD, which can significantly increase its after-pulsing resistance [51]. The after-pulsing probability before launching was less than 0.1%. Based on these studies [51, 57] and our satellite orbit parameters, the expected after-pulsing probability was approximately 0.5% at the end of the two-year mission.

Lastly, we performed a series of space environmental performance and functional tests of the spaceborne, low-noise, Si APD SPDs on the ground, including random vibration, vacuum thermal cycling, and life tests. The results were satisfactory. Typical results are listed in Table 1.

**Table 1. Characteristics of the Spaceborne Low-noise, Si Avalanche Photodiode (APD), Single-photon Detectors (SPDs) in Ground Tests**

| Parameters | Value |
| --- | --- |
| Photon detection efficiency at 780 nm | ~60% |
| Dark count rate | < 0.1 counts per second (cps) at -40 °C |
| Maximum count rate | ~400 K cps |
| Operating temperature | -20 °C to -60 °C |

## 3. In-orbit operations and performance

On 16 August 2016, the spaceborne, low-noise, Si APD SPDs were placed into orbit, onboard the QUESS. Fig. 7(a–c) show the DCRs and operating parameters of the two SPDs, namely SPD #1 and SPD #2, during the 1029 days of in-orbit operations. Operating parameters of temperature (-40 °C) and excess voltages (21 V, 18 V) were first applied to the SPDs to achieve high-detection efficiency and low DCR. As the in-orbit cumulative radiation dose increased, the SPD DCR increased significantly. On the 84th day, the DCR of SPD #1 increased from ~30 cps to ~200 cps, and the SPD operating temperature was then lowered to -50 °C. The excess voltage was slightly adjusted to mitigate the radiation-induced DCR increment. In the first 262 days, the radiation-induced DCRs were kept below 150 cps, and the detection efficiency remained ~60% at the wavelength of 780 nm. During this period, ground–satellite quantum teleportation—the first practical application of the spaceborne, low-noise Si APD SPDs—was demonstrated [15].

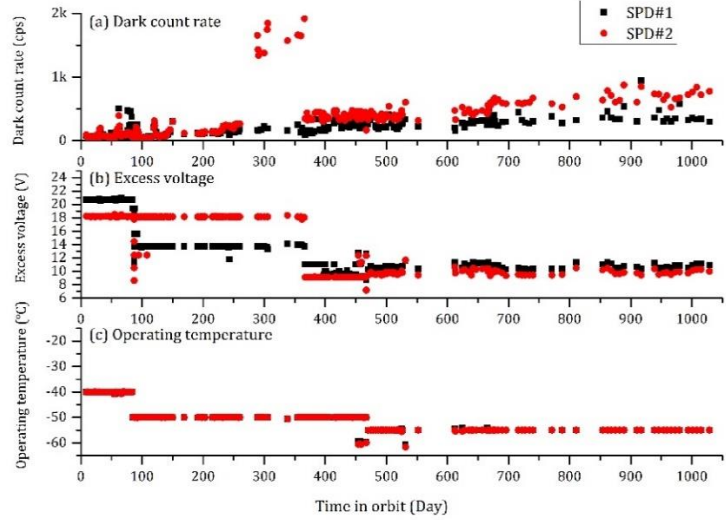

Fig. 7. In-orbit characteristics of SPDs for 1029 consecutive operational days on the QUESS. (a) SPD DCR versus in-orbit time. (b) SPD excess voltage. (c) SPD operating temperature.

On the 289th day, the DCR of the SPD #2 increased abruptly from ~300 cps to ~1600 cps, and deviated from linearity. This phenomenon can be explained according to Eq. (2). When the energy level of the new radiation-induced carrier generation center was very close the intrinsic Fermi energy level, the carrier generation rate of the new radiation-induced carrier generation center would be very high, and would subsequently lead to the sudden DCR increment. Note that the DCR increase was observed in the ground tests when the radiation dose was low, as shown in Fig. 1. To mitigate the sudden radiation-induced DCR increment of SPD #2, a trade-off was established between the DCR and the DE. Optimization of the receiving end signal-to-noise (SNR) rate further lowered the SPD #2 excess voltage. By adjusting the SPD #2, the excess voltage at the cost of a DE of ~15%, we increased the receiving end SNR by approximately four times.

In addition, because the QUESS telescope points at a fixed ground station, the SPD DCRs were unavoidably affected by background noise, especially by the moonlight. This result is shown in Fig. 8. The DCRs fluctuated as a function of the moon phase of the ground-station, and the maximum DCR owing to the moonlight was ~200 cps. It is an effective way to obtain low-background noise by avoiding the effects attributed to the full moon.

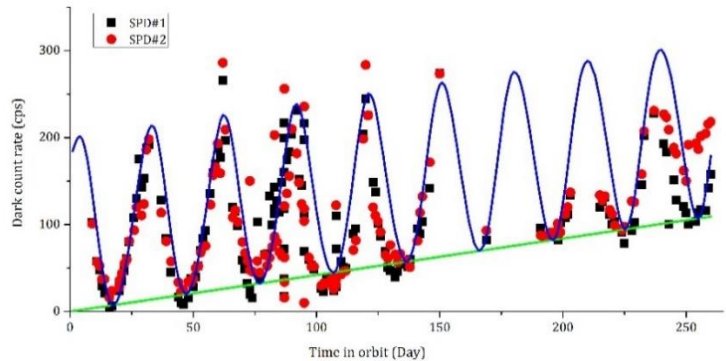

Fig. 8. DCRs of SPD #1 and SPD #2 during the first 262 days in orbit. Black and red dots show the observed data. Green and blue curves represent the radiation-induced DCR and moonlight-induced DCR, respectively. SPD #1 and SPD #2 feature a DCR increase of ~0.56 cps/day during the first 262 days in orbit on the QUESS.

The in-orbit average DCR of SPD #1–2 in the past 1029 days was showed in Fig. 9. By deep cooling and SNR optimization, the spaceborne, low-noise Si APD SPDs have achieved an ultralow, in-orbit DCR increment rate of ~0.54 cps/day, which is two orders of magnitude lower than that of previous results [1, 2], while the DE was maintained above 45%.

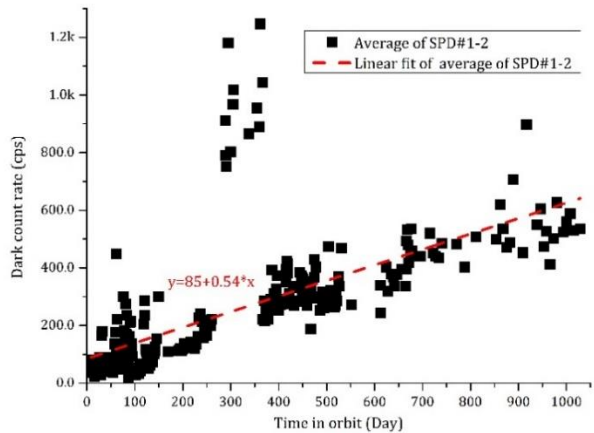

Fig. 9. In-orbit average DCR of SPD #1 and SPD #2 over a 1029-day period.

## 4. Ground–satellite QKD simulations

Spaceborne, low-noise, Si APD SPDs, play an important role in ground-to-satellite or uplink QKD. To generate a secret key, QKD requires the DCR to be kept at a low level to produce a low-quantum bit error rate. Spaceborne, low-noise, Si APD SPDs naturally satisfies this requirement, thus making them suitable for ground-to-satellite QKD. We performed a numerical simulation to study how the DCR affects ground-to-satellite QKD. This involved the adoption of the parameters of an actual satellite-based QKD experiment [58], namely a channel loss of ~40 dB, a pulse generation rate of 100 MHz, a detection efficiency of 60%, and an optical misalignment of 0.5%. The DCRs were the observed data of SPD #1 (see Fig. 7(a)), the conventional increase of ~30 cps/day, and our increase of ~0.56 cps/day. We considered a decoy-state QKD with security against general attacks in the finite-key setting, in which we chose a total number of pulses $N = 10^{11}$ and a security parameter $\varepsilon = 10^{-7}$. The simulation results are shown in Fig. 10.

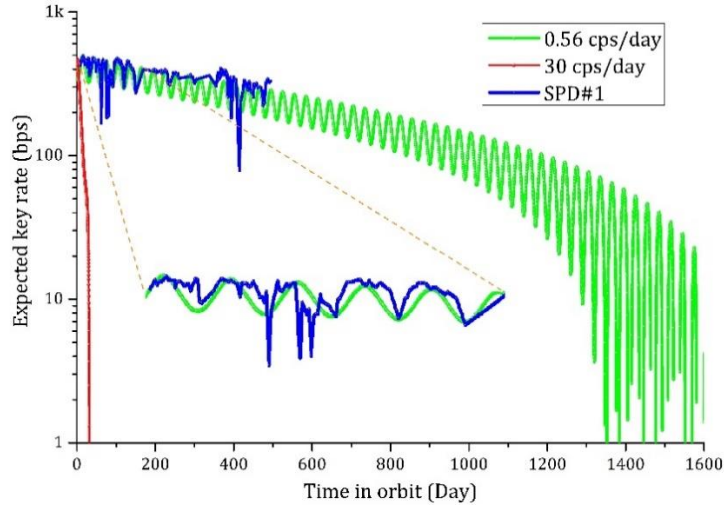

Fig. 10. Expected secret-key rate for ground-to-satellite QKD. The DCRs are (i) those observed in a previous study [15] (blue curve), (ii) those of the spaceborne, low-noise, Si APD SPDs in the present study (green curve), and (iii) those of the previous in-orbit SPDs (red curve). Other simulation parameters are adopted from the satellite-based QKD experiment [15].

We found that our SPDs could enable ground-to-satellite QKD for more than 4.6 years (~1,700 days), whereas a conventional SPD could be used for only one month (~30 days). Besides QKD, the spaceborne, low-noise, Si APD SPD has already enabled the first demonstration of ground–satellite quantum teleportation [15].

## 5. Conclusion

In this study, we have presented the first results from in-orbit tests of spaceborne, low-noise, Si APD SPDs, based on a special shielding structure, multistage cooling technologies, and special driver electronics. By controlling the close-loop DCR, we achieved a low DCR increase of ~0.54 cps/day for 1029 days of in-orbit operations, while we detected efficiencies > 45% at 780 nm. As a potential application in addition to ground–satellite quantum teleportation, ground–satellite QKD was numerically simulated, and results indicated that the spaceborne, low-noise, Si APD SPDs could substantially extend the lifetime of the satellite for QKD.

In future work, we will test the method of annealing [53, 57] in orbit. Moreover, the active quenching method could be applied to the next generation spaceborne, low-noise, Si APD SPDs, and their operation temperature could be further cooled to a lower temperature, such as -100 °C. This may achieve a higher maximum count rate and a lower in-orbit DCR increment. In addition, for the medium and high-orbit satellite, its optoelectronic devices will face high-radiation doses. According to Eq. (4), the APD DCR increment is linearly proportional to the depletion region volume for the same radiation dose. Thus, it is also an effective APD antiradiation measure to use smaller active APD diameters. Furthermore, we have radiated the SPDs with small active diameters, such as the ID Quantique's ID100-SMF20 with an APD of $200 \mu m$, and the

$PD050-CBT from Micro Photon Devices with an APD of $50\mu$m. The radiation test results indicated that the small active diameter of the APD led to an extremely low DCR radiation sensitivity.

Our SPDs provided a practical choice for weak optical signal reception in space applications, such as satellite-based quantum applications, deep-space optical communications, and laser time transfers.

**Funding**